\documentclass[5p,times,twocolumn,number,sort&compress]{elsarticle}

\usepackage{lineno}
\modulolinenumbers[5]

\journal{Journal of \LaTeX\ Templates}

\usepackage{graphicx}  % needed for figures
\usepackage{amsmath}   % for math
\usepackage{amssymb}   % for math
\usepackage{slashed}
\usepackage[caption=false]{subfig}
\usepackage{tabu}
\usepackage[colorlinks=true,
            urlcolor=blue,
            anchorcolor=blue,
            citecolor=blue,
            filecolor=blue,
            linkcolor=blue,
            menucolor=blue,
            linktocpage=true,
            pdfproducer=medialab,
            pdfa=true,
            bookmarks=false]{hyperref}
\usepackage{paralist}
\usepackage{multirow}
\usepackage{etoolbox}
\apptocmd{\thebibliography}{\raggedright}{}{}
\usepackage[capitalize]{cleveref}

\usepackage{capt-of}

\makeatletter
 \def\@textbottom{\vskip \z@ \@plus 4pt}
 \let\@texttop\relax
\makeatother

\begin{document}

\begin{frontmatter}

\title{{\bf\large Measuring CP nature of top-Higgs couplings at the future Large Hadron electron collider}}
%%%%%%%%%%%%%%%%%%%%%%%%%%%%%%%%%%%%%%%%%%%%%%%%%%%%%%%%%%%%%%%%%%%%%%%%%%%%%%%%%%
\author[IITGn]{Baradhwaj Coleppa}
\ead{baradhwaj@iitgn.ac.in}

\author[NITheP]{Mukesh Kumar}
\ead{mukesh.kumar@cern.ch}

\author[IITGn]{Satendra Kumar}
\ead{satendrak@iitgn.ac.in}

\author[Wits]{Bruce Mellado}
\ead{bruce.mellado@wits.ac.za}

\address[IITGn]{Department of Physics, 
	     Indian Institute of Technology Gandhinagar, \\
	     Gandhinagar 382 355, India.}
\address[NITheP]{National Institute for Theoretical Physics,\\
             School of Physics and Mandelstam Institute for Theoretical Physics,\\
             University of the Witwatersrand, Johannesburg, Wits 2050, South Africa.}
\address[Wits]{University of the Witwatersrand, School of Physics,
             Private Bag 3, Wits 2050, South Africa.}

\begin{abstract}
We investigate the sensitivity of top-Higgs coupling by considering the associated vertex as CP phase ($\zeta_t$) dependent 
through the process $p\, e^- \to \bar t \,h \,\nu_e$ in the future Large Hadron electron collider. In particular the decay modes 
are taken to be $h \to b\bar b$ and $\bar t \to$ leptonic mode. 
Several distinct $\zeta_t$ dependent features are demonstrated by considering 
observables like cross sections, top-quark polarisation, rapidity difference between $h$ and $\bar t$ and different angular
asymmetries. Luminosity ($L$) dependent exclusion limits are obtained for $\zeta_t$ by 
considering significance based on fiducial cross sections at different $\sigma$-levels. For electron and proton
beam-energies of 60~GeV and 7~TeV respectively, at $L = 100$~fb$^{-1}$, the regions above $\pi/5 < \zeta_t \leq \pi$ are 
excluded at 2$\sigma$ confidence level, which reflects better sensitivity expected at the Large Hadron Collider. 
With appropriate error fitting methodology we find that the accuracy of SM top-Higgs
coupling could be measured to be $\kappa = 1.00 \pm 0.17 (0.08)$ at $\sqrt{s} = 1.3 (1.8)$~TeV for an ultimate 
$L = 1 \,\rm{ab}^{-1}$. 
\end{abstract}

\begin{keyword}
Electron-Proton collision, top-Higgs coupling, top polarisation
\end{keyword}
%\pacs{13.60.-r, 14.65.Ha, 77.22.Ej, }
\end{frontmatter}

\section{Introduction}
\label{intro}

The recent discovery of the Higgs boson at the Large Hadron Collider (LHC) serves as the last step in establishing 
the particle content of the Standard Model (SM). The next step that has been undertaken is the characterisation 
of its properties regarding spin, CP-nature and the nature of interaction with other particles. While the spin-0 nature of 
the Higgs boson has been established by the experiments~\citep{Aad:2012tfa,Chatrchyan:2012xdj,Aad:2014aba,Khachatryan:2014jba,Aad:2015zhl}  and a complete CP-odd nature excluded at a $99.98$\% confidence limit (C.L.)~\citep{Aad:2015mxa, Khachatryan:2014kca}, the possibility remains that the Higgs boson could still be an admixture of CP-odd and even states. 
Investigation of this possibility  in a future Large Hadron electron Collider (LHeC) is the goal of this article via a detailed 
analysis of the associated production of the Higgs boson with an anti-top quark.

Since in the SM the Higgs boson coupling to fermions is directly proportional to the mass of the fermions, the Yukawa 
coupling associated with the third generation is important in the context of investigating the properties of the Higgs boson. 
Deviations in the top-Higgs coupling directly affects the production cross section of Higgs boson at the colliders, while changes 
in the bottom-Higgs coupling affects the total branching ratios.

Here we study the associated production of the Higgs boson with an anti-top quark at the future $e^-p$ collider 
which employs a $7$~TeV proton beam from a circular $pp$ collider, and electrons from an Energy Recovery Linac 
(ERL) being developed for the LHeC~\citep{AbelleiraFernandez:2012cc,Bruening:2013bga}. 
The choice of an ERL energy of electron of $E_e = 60$ to 120~GeV, with available proton beam energy 
$E_p = 7$~TeV provide centre of mass energy of $\sqrt{s} \approx 1.3$ to 1.8 TeV. 
While the LHC is clearly energetically superior, the LHeC configuration is advantageous for the following reasons:
\begin{inparaenum}[(i)]
\item since initial states are asymmetric, backward and forward scattering can be disentangled,
 \item it provides a clean environment with suppressed backgrounds from strong interaction processes and free from issues
  like pile-ups, multiple interactions etc.
 \item such machines are known for high precision measurements of the dynamical properties of the proton allowing 
 simultaneous tests of electroweak and QCD effects.
\end{inparaenum}
A detailed report on the physics and detector design concepts of the LHeC can be found in the Ref.~\citep{AbelleiraFernandez:2012cc}. A distinguishing feature of the $e^-p$ collider is that the production of the Higgs is only due to electroweak 
processes~\citep{Han:2009pe, Biswal:2012mp} and as noted above, since the $e^-$ and $p$ energies are different, the machine 
can also produce interesting patterns of kinematic distributions that one can exploit to explore the CP nature of the Higgs boson.  

Denoting the  CP-odd (CP-even) components of the top-Higgs coupling by $C_t^P$ ($C_t^S$), the updated bound on the 
CP top-Higgs couplings by combining the LHC Run-1 and Run-2 Higgs data sets allow the ranges $|C_t^P|<0.37$ and $0.85<C_t^S<1.20$, which is  stronger than the previous LHC Run-1 bound $|C_t^S|<0.54$ and $0.68<C_t^S<1.20$. 
We note here that a future precision measurement of the process $e^+e^- \rightarrow h\gamma$ with an accuracy of 
0.5\% will be able to constrain $|C_t^P|<0.19$ at a 240~GeV $e^+e^-$ Higgs factory~\citep{Kobakhidze:2016mfx}. 
Various studies on anomalous top-Higgs coupling in associated production of Higgs and top quark can be
found in~\citep{Cirigliano:2016njn, Cirigliano:2016nyn, Kobakhidze:2014gqa, Li:2017dyz}. 

The article is organised as follows: We discuss the formalism by introducing a generalised CP-phase dependent 
top-Higgs coupling Lagrangian in \cref{lag}. In \cref{ana} simulation and parton-level analyses of the process 
emphasising relevant kinematic observables are discussed. Also in this section we provide luminosity depended exclusion 
limits of phases corresponding to the top-Higgs coupling. Finally, in \cref{conc} we conclude with inferences and 
summary. Though the whole focus of this study is in the LHeC environment, we also discuss and compare our 
results with those expected at the LHC.
\begin{figure}[t]
  \centering
  \subfloat[]{\includegraphics[width=0.15\textwidth]{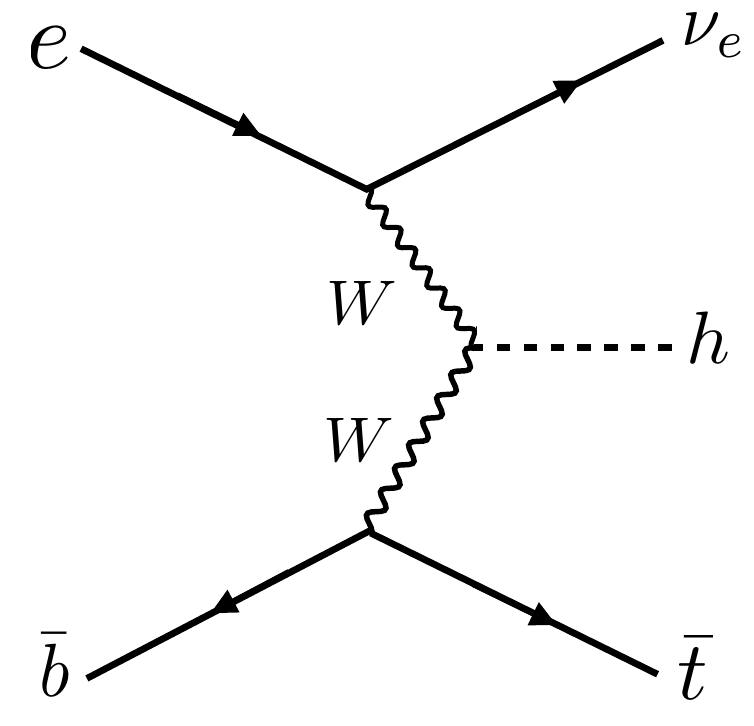}\label{fig:figWa}}
  \subfloat[]{\includegraphics[clip,width=0.15\textwidth]{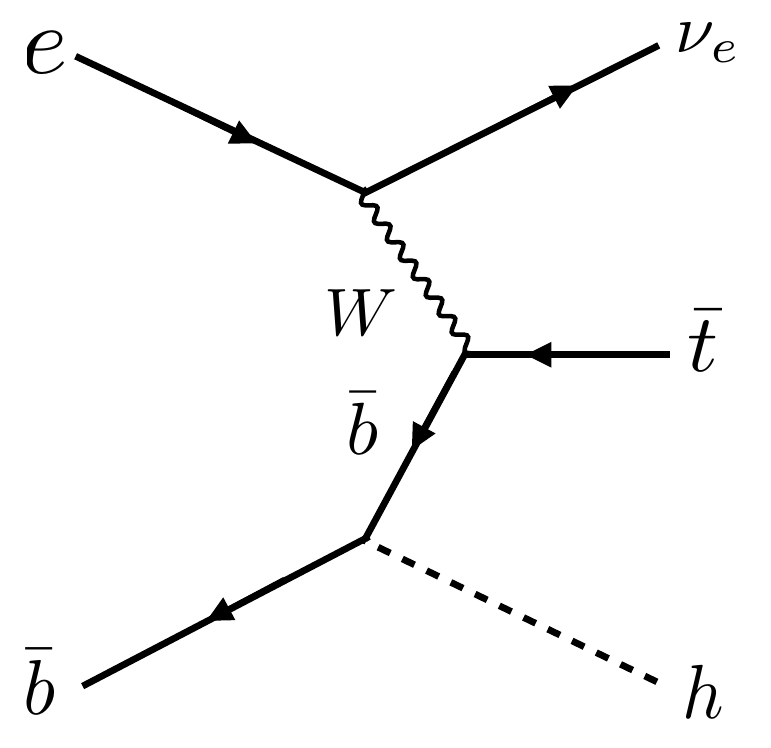}\label{fig:figWb}}
  \subfloat[]{\includegraphics[width=0.16\textwidth]{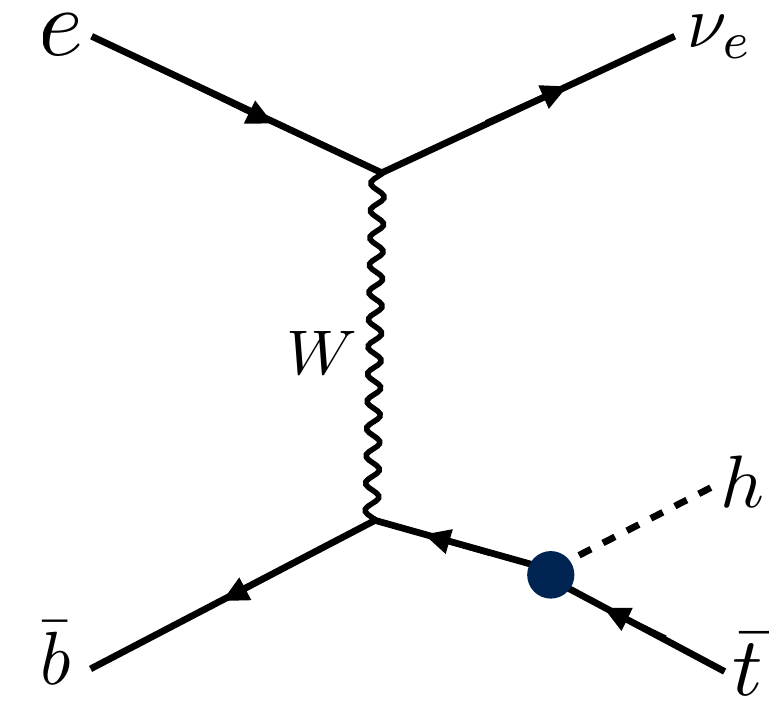}\label{fig:figWc}}
  \caption{\small Leading order Feynman diagrams contributing to the process $p\, e^- \to \bar t\, h\, \nu_{e}$
  at the LHeC. The black dot in the Feynman diagram (c) denotes the top-Higgs coupling which is the subject 
  of this study.}
\label{fig:figW}
\end{figure}
\section{Formalism}
\label{lag}

In the SM, the Yukawa coupling of the third generation of quarks is given by
\begin{equation}
{\cal L}_{\rm Yukawa} = - \frac{m_t}{v} \bar t t h - \frac{m_b}{v} \bar b b h, \label{lyuk}
\end{equation}
where $v\equiv \left(\sqrt{2} G_F \right)^{-1/2} = 2 m_W/g \simeq 246$~GeV, and $m_t$ ($m_b$) 
is the mass of the top (bottom) quark. Due to the pure scalar nature of the Higgs boson in the SM, 
here the top- and bottom-Higgs couplings are completely CP-even. To investigate any beyond the 
SM (BSM) nature of the Higgs-boson as a mixture of CP-even and CP-odd states, we write a CP-phase 
dependent generalised Lagrangian as follows \citep{Rindani:2016scj}: 
\begin{align}
{\cal L}  =& -  \frac{m_t}{v} \bar t~[\kappa \cos\zeta_t+i\gamma_5\sin\zeta_t ]t\,h \notag \\
&-   \frac{m_b}{v} \bar b~[\cos\zeta_b+i\gamma_5\sin\zeta_b ]b\,h. \label{lphase}
\end{align}
Here $\zeta_t$ and $\zeta_b$ are the phases of the top-Higgs and bottom-Higgs couplings respectively. 
It is clear from the Lagrangian in \cref{lphase} that $\zeta_{t, b} = 0$ or $\zeta_{t, b} = \pi$ correspond to 
a pure scalar state while $\zeta_{t, b} =\frac{\pi}{2}$ to a pure pseudo scalar state. Thus, the ranges 
$0<\zeta_{t, b}<\pi/2$ or $\pi/2<\zeta_{t, b}<\pi$ represent a mixture of the different CP-states. The 
case $\kappa = 1$, $\zeta_t = 0$ corresponds to the SM. 
In terms of $C^S_t$ and $C^P_t$, we can also translate $\zeta_t = \tan^{-1}(C_t^P/C_t^S)$.

At the LHeC, the top-Higgs couplings can be probed via associated production of Higgs-boson with 
anti-top quark $p \,e^- \to \bar t \,h\,\nu_e$ - it is thus necessary to consider a 5-flavour proton including the $b$-quark 
parton distribution. The Feynman diagrams for the process under investigation are shown in \cref{fig:figW}. 
It is important to notice that in this process three important couplings are involved, namely $hWW$, $Wtb$ and 
the top-Higgs ($tth$). A detailed study of $hWW$ and $Wtb$ couplings at the $e^- p$ collider have been 
performed in Refs.~\citep{Biswal:2012mp,Kumar:2015kca} and~\citep{Dutta:2013mva}, respectively.
For our studies we do not consider the BSM bottom-Higgs coupling since the effect of the phase $\zeta_b$ 
on the total production cross section or kinematics of top-Higgs production at the LHeC are negligible. 
Thus in what follows, we simply set $\zeta_b = 0$.

As noted in Ref.~\citep{Rindani:2016scj} in the context of the LHC, quantitatively an interesting feature 
can be observed: in the pure SM case there is constructive interference between the diagrams shown in 
\cref{fig:figWa} and \cref{fig:figWc} for $\zeta_t >\pi/2$ resulting in an enhancement in 
the total production cross section of associated top-Higgs significantly. This is also true for $\zeta_t < \pi/2$ - however 
the degree of enhancement is much smaller owing to the flipped sign of the CP-even part of the coupling.
\begin{figure}[t]
  \includegraphics[width=0.48\textwidth,height=0.30\textwidth]{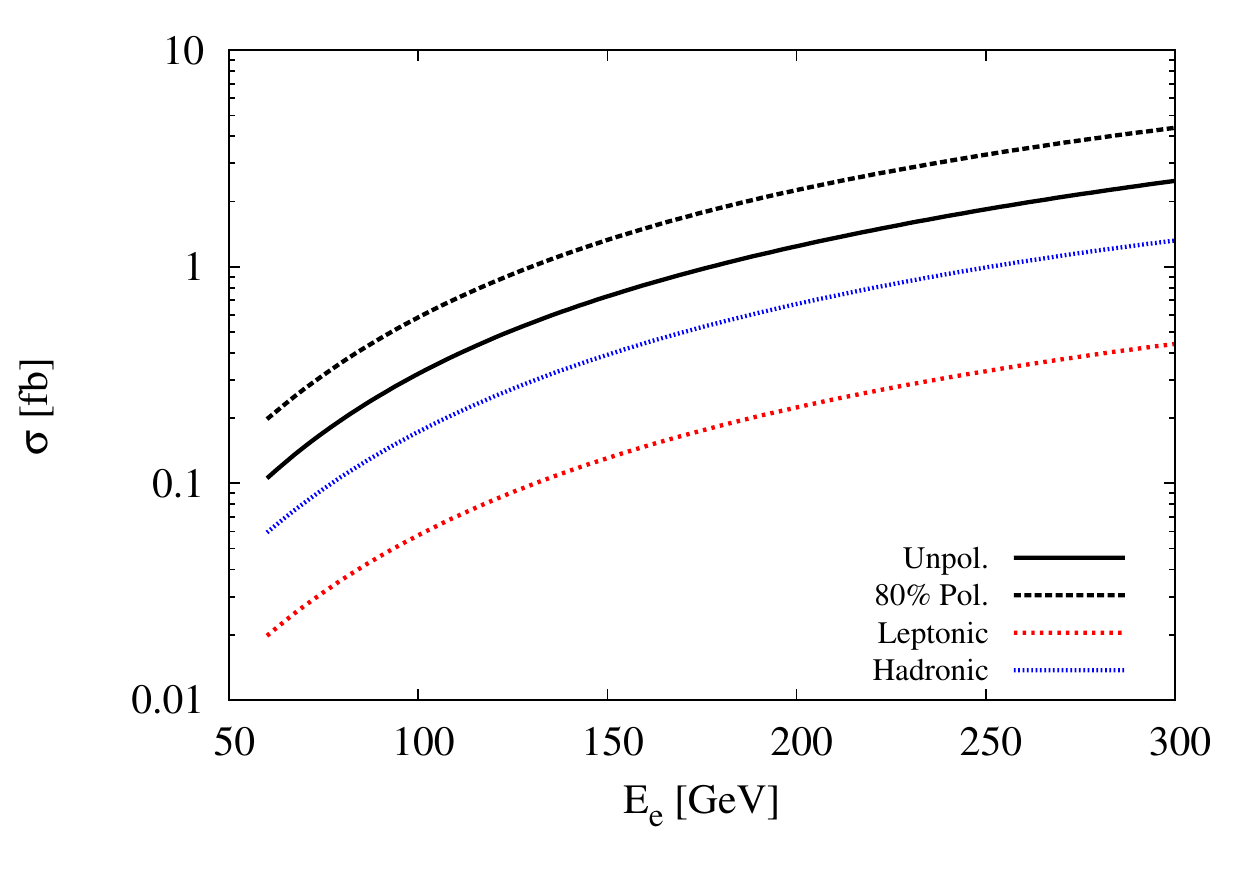}
  \caption{\small Total cross section of the associated top-Higgs production against electron beam energy 
  for fixed $E_p =$~7 TeV. The dotted and solid $black$ lines correspond to the process 
  $p~ e^- \rightarrow~ \bar t~ h~ \nu_{e}$ with and without polarisation of electron beam respectively. 
  The dotted $red$ and $blue$ lines correspond to $\sigma\times$BR for the leptonic and hadronic decay modes of 
  $\bar{t}$  where for this estimation we use basic cuts as given in text.}
\label{fig:cs_rs}
\end{figure}
\section{Simulation and analysis}
\label{ana}

We begin our study to probe the sensitivity of the top-Higgs couplings in terms of $\zeta_t$ by building a model
file for the Lagrangian in \cref{lphase} using \texttt{FeynRules}~\citep{Alloul:2013bka}, and then simulating the charged 
current associated top-Higgs production channel $p\, e^- \to \bar t \, h\, \nu_e$ (see \cref{fig:figW}), with $h$ further 
decaying into a $b \bar b$ pair and the $\bar t$ decaying leptonically in the LHeC set-up with centre of mass energy
of $\sqrt{s} \approx 1.3$~TeV.  
In this article we perform the analysis at parton level only where for signal and background event generation we 
use the Monte Carlo event generator package \texttt{MadGraph5}~\citep{Alwall:2014hca}. 
We use \texttt{NN23LO1}~\citep{Ball:2012cx, Deans:2013mha} 
parton distribution functions for all event generations. 
The factorisation and renormalisation scales for the signal simulation are fixed at $\mu_F = \mu_R = (m_t + m_h)/4$ 
while background simulations are done with the default \texttt{MadGraph5}~\citep{Alwall:2014hca} dynamic scales. 
The $e^-$ polarisation is assumed to be $- 80$\%.
We now list and explain various kinematic observables that can serve as possible discriminants of a CP-odd $t\bar{t}h$ 
coupling.

\subsection{Cross section studies}
\label{cs}
\begin{table}[t]
\centering
\resizebox{\linewidth}{!}{
{\tabulinesep=5pt
\begin{tabu}{lccc}\hline
Process                        & {\scshape cc} (fb)    & {\scshape nc} (fb)    & {\scshape photo} (fb)  \\ \hline
Signal:                          & $1.98 \times 10^{-2}$ & $--$ & $--$ \\ 
$Wjjj+X$,  $\backslash h$     & $2.05 \times 10^{+2}$ & $3.18 \times 10^{+1}$ & $3.40 \times 10^{+3}$ \\ 
$Wjjj+X$,  $\backslash t$       & $4.18 \times 10^{+1}$ & $3.16 \times 10^{+1}$ & $3.41 \times 10^{+3}$ \\ 
$Wjjj+X$,  $\backslash t h$    & $4.16 \times 10^{+1}$ & $3.18 \times 10^{+1}$ & $3.41 \times 10^{+3}$ \\ \hline
\end{tabu}}
}
\caption{\small Cross sections of signal and backgrounds in charged current
                ({\scshape cc}), neutral current ({\scshape nc}) and photo-production ({\scshape photo})
                modes for $E_e = 60$ GeV and $E_p = 7$ TeV as explained in the text. 
                Here $X$ could be either of missing energy
                or electron and $j $ is all possible combinations of light-, $c$- and $b$-quarks and gluons.
                For this estimation we use basic cuts as mentioned in text and electron polarisation is 
                taken to be~$-\,0.8$.}
\label{tab:xsec}
\end{table}
In \cref{fig:cs_rs}, we present the variation of the total cross section against the electron 
beam energy for the signal process $ p\,e^- \to \bar t h \nu_e$, by considering un-polarised and~$-\,80$\% polarised 
$e^-$ beam. Also, the effect of branchings of $h \to b \bar b$ and the $\bar{t}$ decay for both leptonic and hadronic 
modes are shown.    
Possible background events typically arise from $W$+ multi-jet events, $Wb\bar b\bar b$ with missing energy which 
comes by considering only top-line ($\backslash h $), only Higgs-line ($\backslash t $) and without top- and 
Higgs-line ($\backslash th $) in charged and neutral current deep-inelastic scattering and in photo-production 
by further decaying $W$ into leptonic mode.
In \cref{tab:xsec} we have give an estimation of cross sections for signal and all possible backgrounds imposing only 
basic cuts on rapidity $|\eta| \leq 10$ for light-jets, leptons and $b$-tagged jets, the transverse momentum 
cut $p_T \geq 10$~GeV and $\Delta R_{\rm min}$\footnote{The distance parameter between any two particles is defined 
as $\Delta R = \sqrt{(\Delta \phi)^2 + (\Delta \eta)^2}$, where $\phi$ and $\eta$ are the azimuthal angle and 
rapidity respectively of particles into consideration.} = 0.4 for all particles.

\begin{figure}[t]
\begin{center}
\begin{tabular}{c}
\includegraphics[angle=0,width=80mm]{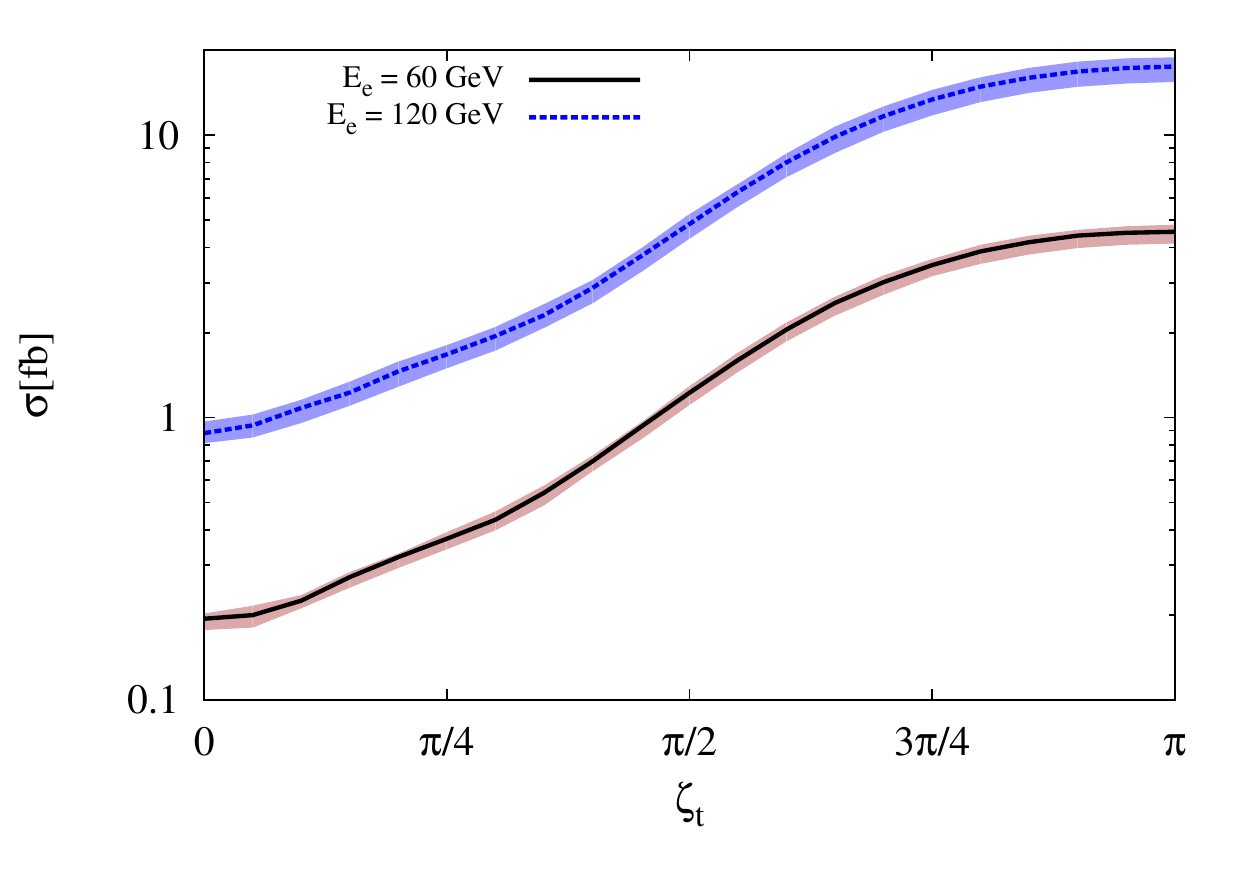}
\end{tabular}
\vspace{-4mm}
\caption{\small {Total cross section as a function of $\zeta_t$ with scale uncertainties. The $black$ solid and $blue$ dotted 
lines correspond to $E_e = 60$ and $120$~GeV respectively for fixed $E_p = 7$~TeV and $\mu_F = \mu_R = (m_t + m_h)/4$.}}
\label{fig:mu}
\end{center}
\end{figure}

We now estimate the sensitivity of the associated top-Higgs production cross-section, $\sigma(\zeta_t)$, 
as a function of the CP phase of the $tth$-coupling as shown in \cref{fig:mu} by considering $E_e = 60$ 
and $120$~GeV with fixed $E_p = 7$~TeV. 
The scale uncertainties are taken as $(m_t + m_h)/8 \leq \mu_F = \mu_R \leq (m_t + m_h)/2$. 
Here $\sigma(\zeta_t = 0)$ corresponds to the SM cross section.
We notice that the cross section is very sensitive to $\zeta_t$ in the region 
$\zeta_t > \frac{\pi}{2}$ where the interference between the diagrams becomes constructive. 
Below $\zeta_t = \frac{\pi}{2}$ the interference is still constructive though its degree decreases with 
$\zeta_t$, thus increasing the cross section by around 500\% at $\zeta_t = \frac{\pi}{2}$ which corresponds 
to the pure CP-odd case. 
On the other hand, for pure CP-even case $\zeta_t = \pi$ with opposite-sign of $tth$-coupling the cross section 
can be enhanced by up to 2400\% for $E_e = 60$~GeV. Notice that for the case $E_e = 120$~GeV, $\sigma(\zeta_t)$ 
displays a similar shape with enhanced cross sections with respect to $E_e = 60$~GeV case.
The scale uncertainty on an average is approximately 7(9)\% for $E_e = 60 (120)$~GeV in the whole range
of $\zeta_t$.

However, it is quite interesting that the combined ATLAS and CMS measurements at $\sqrt{s} = 7$ and $8$~TeV
allow deviation of cross section in terms of signal strength $\mu = 2.3^{+0.7}_{-0.6}$~\citep{Khachatryan:2016vau} 
for associated top-Higgs production\footnote{Note that at the LHC the production of associated Higgs boson with 
top-quark is possible via double and single-top quarks and is different from LHeC where the environment and 
centre of mass energies are different. The signal strength is defined as 
$\mu = \sigma_{\rm observed}/\sigma_{\rm SM}$.}. Though one may investigate the possibilities of such observations
due to comparatively heavy scalar with respect to the Higgs-boson as in 
Ref.~\citep{vonBuddenbrock:2015ema, vonBuddenbrock:2016rmr}.

\subsection{Rapidity difference between the anti-top and the Higgs}
\label{deltay}
In Refs.~\citep{Biswal:2012mp,Kumar:2015kca} it was suggested that in order to explore the tensorial spin-CP 
nature of $hW^+W^-$ and $hhW^+W^-$ vertices, azimuthal angle correlation between missing energy and scattered 
jets are a good observable. Also further studying the asymmetry based on such observables proves to be an excellent tool
for any BSM nature of the associated couplings. Here and in the next subsections we include such observables in our 
studies with different combinations of final state particles as a function of $\zeta_t$. 
We begin with the sensitivity of BSM aspects of the $tth$ coupling in the rapidity difference between the anti-top 
quark and the Higgs boson distribution, $\Delta y_{\rm ht}$.     

\begin{figure}[t]
\begin{center}
\begin{tabular}{c}
\includegraphics[angle=0,width=80mm]{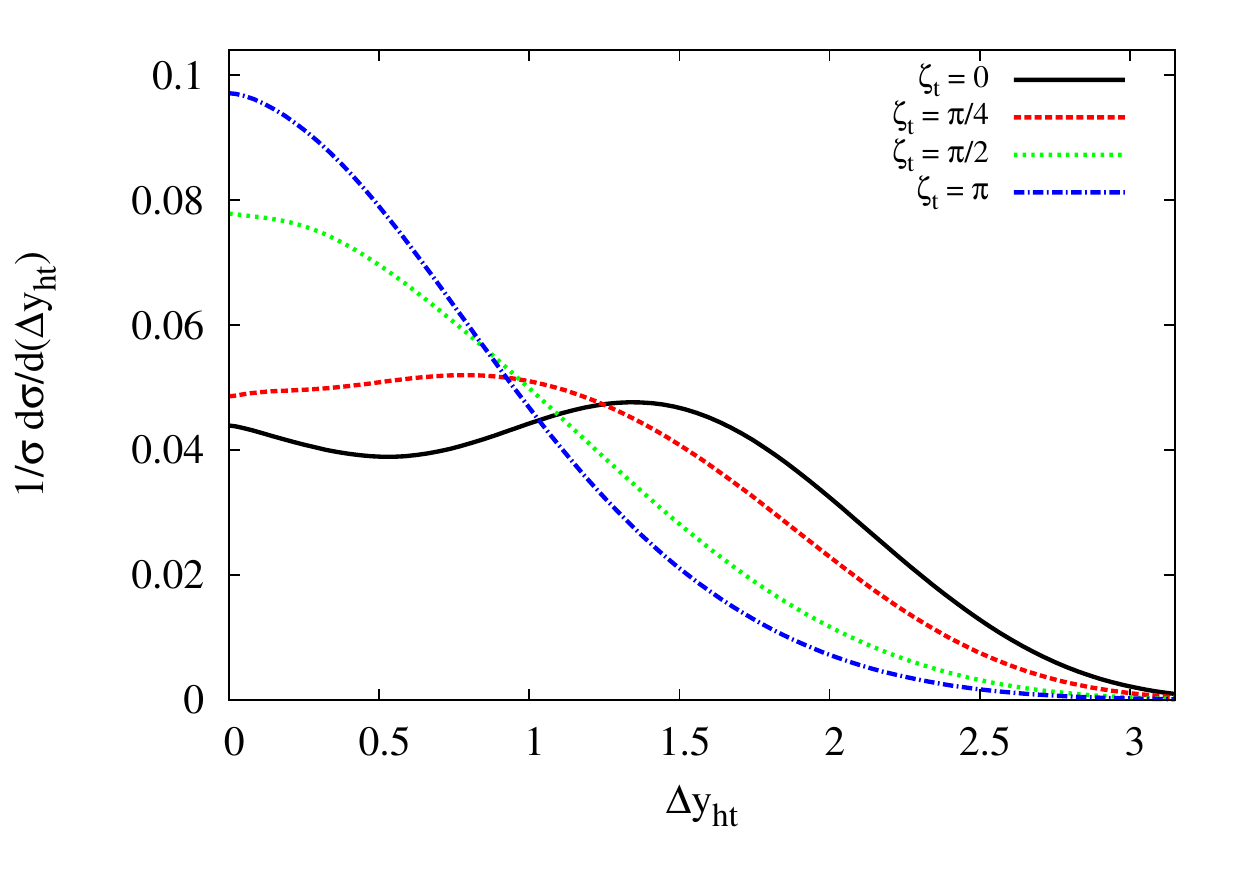} 
\end{tabular}
\vspace{-4mm}
\caption{\small {The normalised difference between rapidities of top quark and the Higgs boson at some 
typical values of $\zeta_t$ for $E_e = 60$~GeV and $E_p = 7$~TeV. The $black$ solid line corresponds to 
the SM case, while dotted lines correspond to different values of $\zeta_t$.}}
\label{fig:deltayht}
\end{center}
\end{figure}

In Fig.~\ref{fig:deltayht} we present the normalised $\Delta y_{\rm ht}$ distribution for a few chosen 
values of $\zeta_t$.  
Any BSM physics effect can be observed by comparing the shape corresponding to the SM case $\zeta_t=0$. 
We find that the distribution features for the different values of CP phase split 
into two distinguishable regions when $\Delta y_{\rm ht} < 1$ and $1 < \Delta y_{\rm ht} < 3$. In the former, most 
values of $\zeta_t$  are seen to correspond to distributions larger than the SM case, while the second region 
presents a complementary behaviour. The distortion in the shape for $\zeta_t > 0$ is the effect of mixing between
CP-even and odd components of the $tth$ vertex following the Lagrangian in \cref{lphase}. 
Overall, with the inclusion of spin-0$^+$ BSM admixture, the $\Delta y_{\rm ht}$ distribution is pushed towards
lower values and act as a potential discriminator to explore the CP-nature of $tth$-coupling.  
Similar studies are used to probe the tensor structure of $hVV$ ($V = W^\pm, Z$) coupling at the LHC and one 
such study of the Higgs boson production in the vector boson fusion mode is performed in~\citep{Kruse:2014pya} 
by taking the rapidity difference between the Higgs
and the leading parton. 

\subsection{Top quark polarisation}
\label{top_pol}
The large top-quark mass $m_t = 172.84 \pm 0.70$~GeV~\citep{Aaboud:2016igd} indicates that the top could potentially
play a singular role in the understanding of electroweak symmetry breaking in BSM scenarios.   
Since the decay width of the top exceeds $\Lambda_{\textrm{QCD}}$, the top decays before hadronising and thus its spin 
information is preserved in the differential distribution of its decay products. With the Higgs coupling to top modified, it is 
reasonable to expect an asymmetry in the production of tops of different polarisations and the effect of $\zeta_t$ on this 
asymmetry. 

\begin{figure}[t]
\begin{center}
\begin{tabular}{c}
\includegraphics[angle=0,width=80mm]{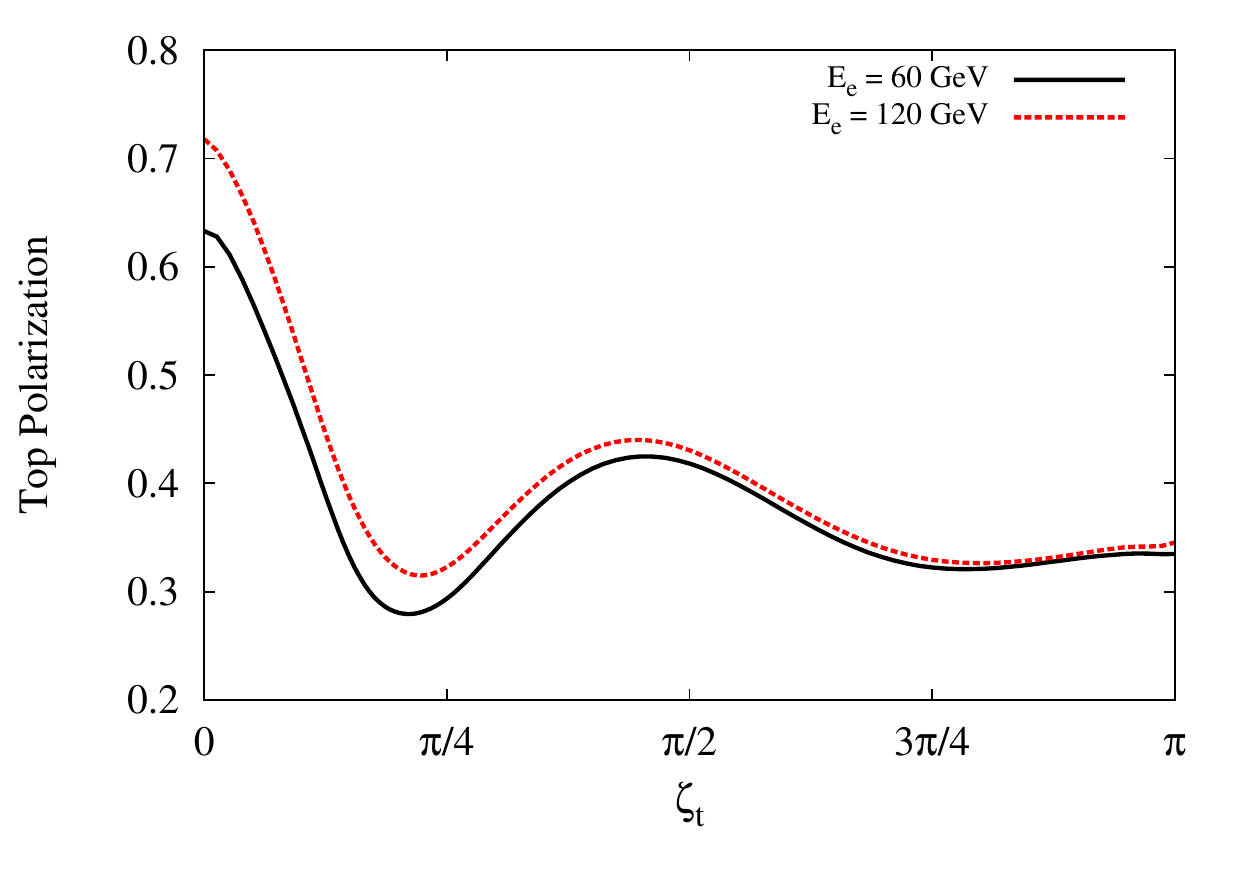} 
\end{tabular}
\vspace{-4mm}
\caption{\small{The degree of longitudinal polarisation ($P_t$) of the top quark against $\zeta_t$. 
The $black$ solid and $red$ dotted lines correspond to the $E_e = 60$ and $120$~GeV, 
while $E_p$ is fixed at 7~TeV.}}
\label{fig:pol}
\end{center}
\end{figure}

We define the degree of longitudinal polarisation $P_t$ of the top quark as
\begin{align}
P_t = \frac{N_+~-~N_-}{N_+~+~N_-} \equiv \frac{\sigma_+~-~\sigma_-}{\sigma_+~+~\sigma_-},
\label{eqn:top_pol}
\end{align}
where $N_+$ and $N_-$ denote the number of events with positive and negative helicity anti-top quarks
respectively, which can be rewritten in terms of the corresponding cross sections $\sigma_\pm$. 
In \cref{fig:pol}, we present $P_t$  in the process $p\,e^-\to \bar t\, h\, \nu_e$ at the LHeC as a 
function of $\zeta_t$. We obtain $N_\pm$ or $\sigma_\pm$ using the helicity amplitudes in \texttt{MadGraph5}.
It can be seen from the plot that the degree of polarisation is quite sensitive over the entire range of $\zeta_t$ since
the CP-odd coupling violates parity for any non-zero $\zeta_t$.
It is interesting to note that if \cref{fig:figWc} is the only diagram that contributed to $P_t$ then the fraction of 
right-handedly polarized anti-top quark would increase as $\zeta_t$ increases from $0$ and reach a maximum at 
$\zeta_t=\pi/2$ and then fall. 
However, the presence of other diagrams means that the plot is not symmetric about $\zeta_t=\pi/2$.  The general 
features of $P_t$ in \cref{fig:pol} can be understood as the effect of interference among the diagrams 
in \cref{fig:figWa}, \cref{fig:figWb} (from where right-handed anti-top quarks are produced) and the Higgs-bremsstrahlung 
diagram \cref{fig:figWc}, which contains the CP-violating $\sin\zeta_t$ term.
   
As mentioned before, information of the spin of the top is preserved in its decay products
and the angular distribution of its decay products can be parametrised as:
\begin{align}
\frac{1}{\Gamma_f} \frac{d \Gamma_f}{d\cos\theta_f} =\frac{1}{2}(1+\alpha_f P_t \cos\theta_f),
\label{ang_decay}
\end{align}
where $f$ is the type of top decay product,
$\theta_f$ is the angle between $f$ and the top-quark spin quantisation axis measured in the rest 
frame of the top-quark and  $\Gamma_f$ denotes the partial decay width corresponding to $f$. 
For the decay mode $t \to b + W^\pm (\to l^\pm + \nu_l)$
at lowest order, $\alpha_W = -\alpha_b = 0.39, \alpha_\nu=-0.3, \alpha_l=1$ \cite{Bernreuther:2008ju}, 
with small QCD corrections to these values \cite{Czarnecki:1990pe, Brandenburg:2002xr}. 
The charged lepton $l^\pm$ (or the down-type quark $d$ in a hadronic decay of the intermediate W) is
nearly 100\% correlated with the top quark spin which means that the  $l^\pm$ or 
$d$ is much more likely to be emitted in the direction of the top quark spin than in the opposite direction.
It is a well known fact that the energy and momentum of leptons can be measured with high precision at the LHC
and the same is true for the LHeC as well, so we focus on the leptonic decay mode of the anti-top for 
asymmetries in angular observable studies in what follows.
\subsection{Cut-based event optimisation}
\label{optim}
Before discussing the angular observables for this study, it is important to discuss the optimisation of
SM signal and background events as mentioned in \cref{cs}. 
Angular observables are affected due to kinematic cuts
and hence it is better to analyse events after optimising the signal with respect to backgrounds. 
The full SM signal process for this analysis is $p\,e^- \to \bar t \,h \,\nu_e$, with
$h \to b \bar b$ and $\bar t \to W^- \bar b, W^- \to l^- \nu_l$  ($l^\pm = e^\pm, \mu^\pm$).
After preliminary analysis of various kinematic distributions of final state particles of the SM
signal and all possible leptonic backgrounds, we employ the following criteria to select events:
\begin{inparaenum}[(i)]
 \item $p_T \ge 20$~GeV for $b$-tagged jets and light-jets, and $p_T \ge 10$~GeV for leptons.
 \item Since the LHeC collider is asymmetric, event statistics of final state particles are mostly 
 accumulated on the left or right sides of the transverse plane $\eta = 0$ (depending on the initial direction 
 of $p$ and $e^-$) - we select events within $-2 \le \eta \le 5$ for $b$-tagged jets while $2 \le \eta \le 5$ for  
 leptons and light-jets,
 \item The separation distance of all final state particles are taken to be $\Delta R > 0.4$.
 \item Missing transverse energy $\slashed{E}_T > 10$~GeV to select the top events.
 \item Invariant mass windows for the Higgs through $b$-tagged jets and the top
 are required to be $115 < m_{bb} < 130$~GeV and $160 < m_t < 177$~GeV respectively, which are 
 important to reduce the background events substantially.
 \end{inparaenum}
 In these selections the $b$-tagging efficiency is assumed to be 70\%, with fake rates from $c$-initiated jets
 and light jets to the $b$-jets to be 10\% and 1\% respectively. 
These constitute our event selection criteria which we use in the subsequent analysis.
  
There are two major difficulties in reconstructing the Higgs boson and the top in the process 
$p \, e^- \to \bar t \, h \, \nu_e$ $\to (W^- \bar b) h \nu_e$ $\to l^- \nu_l \bar b b \bar b \nu_e$:
\begin{inparaenum}[(a)]
\item Choosing appropriate $b$-tagged jets - in the final state we have 3 $b$-tagged jets with two originating from 
$h$ decay and one from the decay of $\bar{t}$ and 
\item The source of missing energy comes from both the production process and from $W^\pm$ decay. 
\end{inparaenum}
 Since we performed parton-level analysis, we read the event files 
 generated from the Monte Carlo generator and by reading appropriate identities we obtained information about the 
 origin of $b$-tagged jets and neutrino and the corresponding four-momenta information was used for the analysis. 
 Although the detector-level analysis is beyond the scope of this article, we mention briefly that for distinguishability 
 of $b$-jets the solution is to take into account the $p_T$ ordering of all $b$-tagged jets and since top-quark is heavier 
 than the Higgs boson, the leading-$p_T$ $b$-jet can identified as the decay product of top-quark, and the sub-leading 
 and next to sub-leading $p_T$-ordered $b$-jets can be used to reconstruct Higgs boson. 
 
 To reconstruct the top, substantial requirement on missing energy and top-quark invariant mass formula
 $m_t^2 = (m_T + m_{b_1})^2$ can be used, where $m_T$ is transverse mass observable to reconstruct 
 $W$-boson and $m_{b_1}$ is the mass of leading-$b$ jet and is given as:
 \begin{align} 
 m_T = \sqrt{2\, p_T^{l}\, p_T^{\nu}\, (1 - \cos (\phi_{l} - \phi_\nu))}, \notag
 \end{align} 
 where $\cos (\phi_l - \phi_\nu )$ is the angle between the electron and neutrino in the transverse plane,
 and $\phi_l$ ($\phi_\nu$) is the azimuthal angle of the electron (neutrino). 
 However, it is to be noted that $m_T$ is also inefficient when there are more than one sources of missing energy and 
 hence alternative method should be explored.
\begin{figure}[t]
\begin{center}
\begin{tabular}{c}
\hspace{-6mm}
\includegraphics[angle=0,width=88mm]{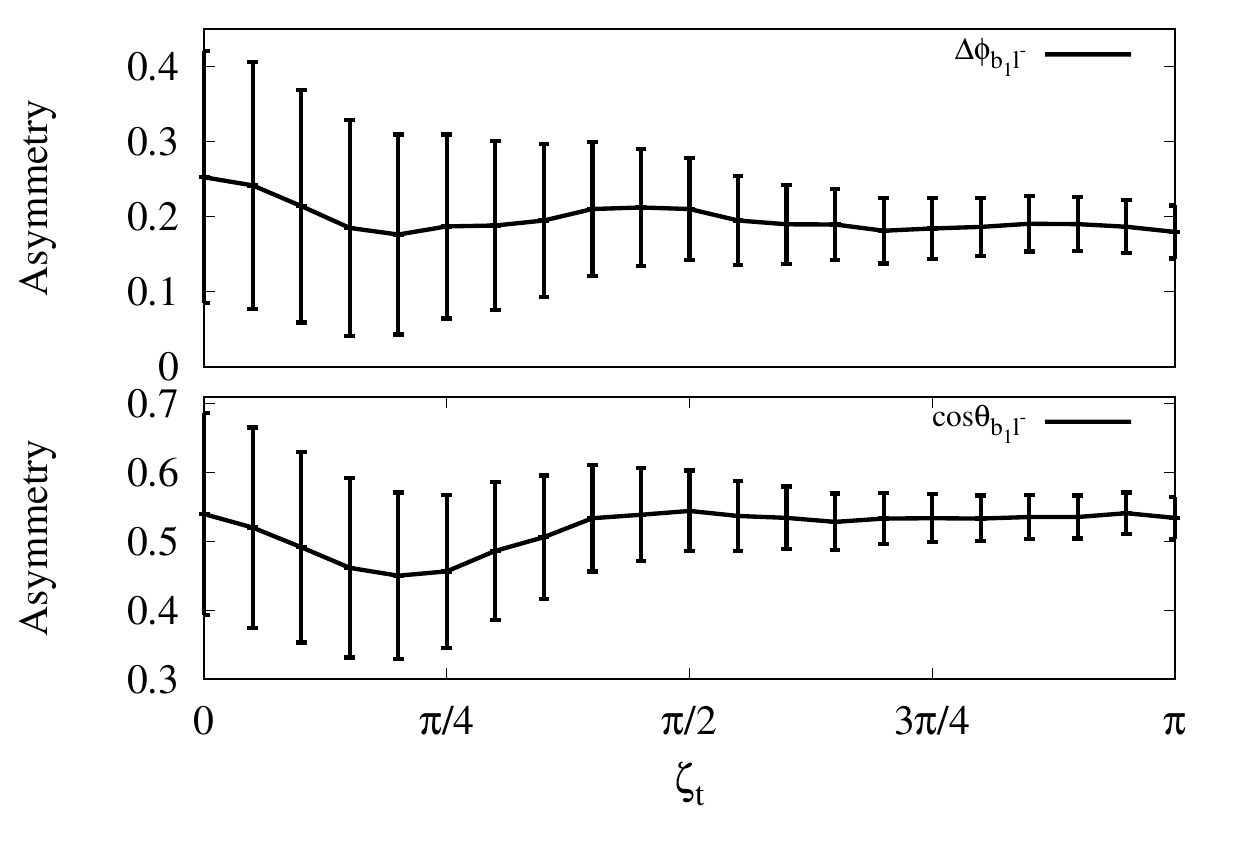}
\end{tabular}
\vspace{-4mm}
\caption{\small{Variation of angular asymmetries between the leading $b$-tagged jet and the charged lepton 
in the differential azimuthal and polar angle ($\Delta\phi_{b_1l^-}$ and $\cos\theta_{b_1l^-}$) distributions 
with respect to $\zeta_t$ for $E_e =60$~GeV and $E_p = 7$~TeV. The error bars
correspond to the uncertainties in asymmetry measurement at $L =  1 \,{\rm ab}^{-1}$.}}
\label{fig:asymmt}
\end{center}
\end{figure}
          
\subsection{Angular observables in terms of asymmetries}
\label{asymm}

After this short discussions on event selection criteria, we now discuss observables based on angular 
asymmetry between different final state particles.
We construct the asymmetry from the differential distribution of kinematic observables using the final leptons and 
$b$-tagged jets. These asymmetries are studied only for signal processes as a function of $\zeta_t$. 
The angular asymmetries with respect to polar angle\footnote{Polar angle $\cos\theta_{ij}(p_i,p_j)$ between two 
final state particles $i$ and $j$ with four-momentum $p_i$ and $p_j$ respectively is defined as the angle between 
direction of $p_i$ in the rest frame of $p_i + p_j$ and the direction of $p_i + p_j$ in the lab frame.}
$\cos\theta_{ij}$ and the azimuthal angle difference $\Delta\phi_{ij}$ are defined to be:
\begin{align}
A_{\theta_{ij}} =
\frac{N_+^A(\cos\theta_{ij} > 0) - N_-^A(\cos\theta_{ij} < 0)}{N_+^A(\cos\theta_{ij} > 0) + N_-^A(\cos\theta_{ij} < 0)},
\label{fig:asy1}
\end{align}
\begin{align}
A_{\Delta\phi_{ij}} = 
\frac{N_+^A(\Delta\phi_{ij} > \pi/2) - N_-^A(\Delta\phi_{ij} < \pi/2)}{N_+^A(\Delta\phi_{ij} > \pi/2) + N_-^A(\Delta\phi_{ij} < \pi/2)},
\label{fig:asy2}
\end{align}
where $i$ and $j$ are any two different final state particles.
Using binomial distribution we use the following formula to calculate the statistical uncertainty ($\delta_\alpha$) in the 
measurement of these asymmetries ($A_{\alpha}$):
\begin{equation}
\delta_\alpha = \sqrt{\frac{1-A^2_{\alpha} (\zeta_t)}{\sigma_{\zeta_t} \cdot L}}, \qquad \qquad(\alpha = \theta_{ij}, \Delta\phi_{ij})
\label{error}
\end{equation}
where $\sigma_{\zeta_t}$ is the total cross section of signal events as a function of $\zeta_t$ and $L$ 
is the total integrated luminosity. 

In \cref{fig:asymmt}, we show the asymmetries between the charged lepton and the $\bar{b}$ from $\bar{t}$ decay 
(denoted by $b_1$ in the plot) as functions of $\zeta_t$. We can see that the asymmetries in 
$\Delta\phi_{b_1\ell^-}$ and $\cos\theta_{b_1\ell^-}$ follow the top polarisation curve to some extent in that they fall 
till $\zeta_t\approx\pi/4$. 
We find that beyond $\zeta_t=\pi/2$, the curves flatten.  As explained in the \cref{top_pol} the shape in these asymmetry 
observables are also influenced by interference among the Feynman diagrams shown in \cref{fig:figW}.    
Overall we can conclude that these asymmetry observables can serve as good discriminators for a non-zero $\zeta_t$, 
particularly for $\zeta_t<\pi/2$ where the difference from the $\zeta_t=0$ case is more pronounced. 
%%%%%%%%%%%%%%%%%%%%%%%%%%%%%%%%%%%
\subsection{Exclusion limits}
\label{excl}
\begin{figure}[t]
\begin{center}
\begin{tabular}{c}
\includegraphics[angle=0,width=82mm]{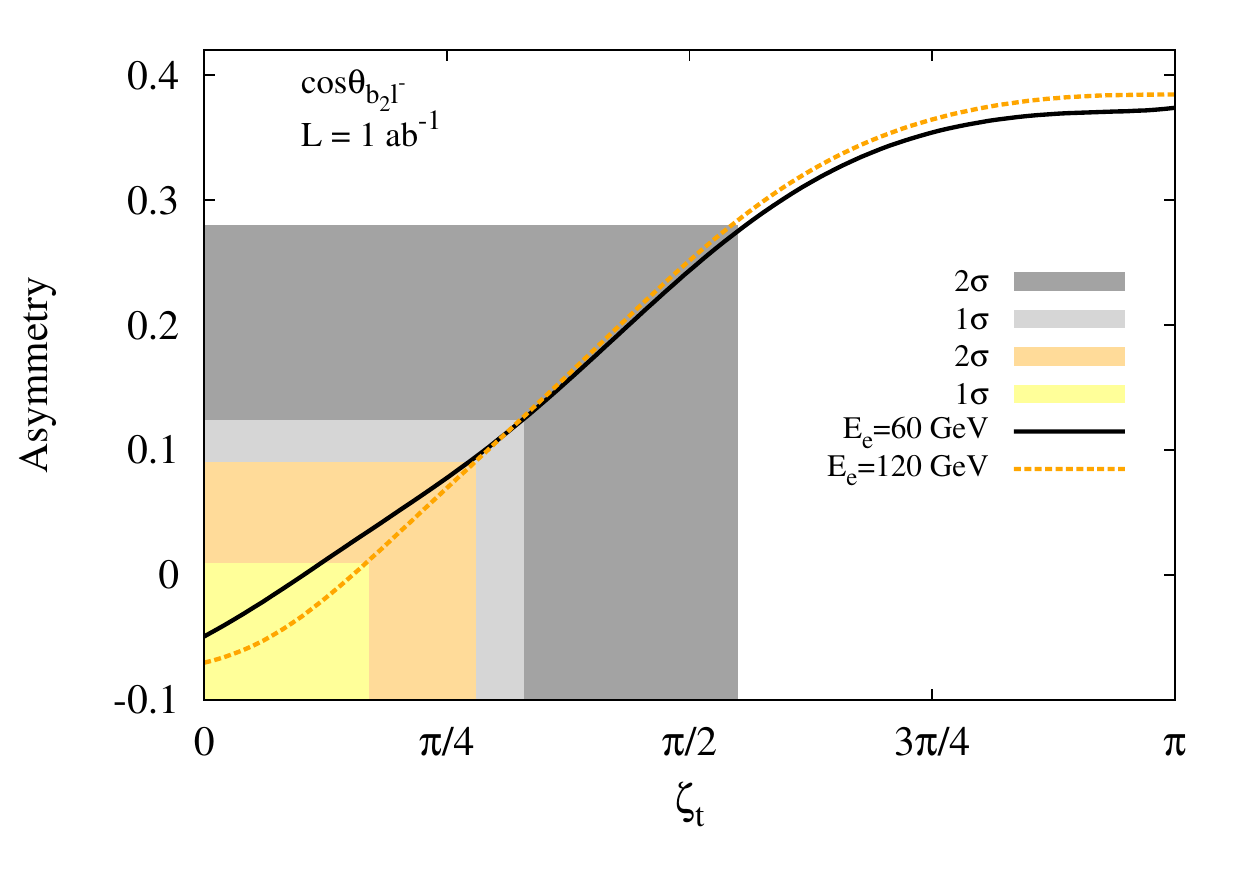}\\
\end{tabular}
\vspace{-4mm}
\caption{\small{Variation of the angular asymmetry between the subleading $b$-tagged jets and charged leptons 
in the differential polar angle ($\cos\theta_{b_2l^-}$) distribution with respect to $\zeta_t$ for $E_e = 60$~GeV 
($black$ solid line) and $E_e = 120$~GeV ($orange$ dashed line) with $E_p = 7$~TeV. The shaded regions
$grey$ ($orange$) and $light~grey$ ($yellow$) corresponds to $2\sigma$ and $1\sigma$ of statistical uncertainty in the 
measurement of the asymmetry in the SM for $E_e = 60 \,(120)$~GeV at $L =  1\,{\rm ab}^{-1}$ respectively.}}
\label{fig:Excl1}
\end{center}
\end{figure}
\begin{figure}[t]
\begin{center}
\begin{tabular}{c}
\includegraphics[angle=0,width=85mm]{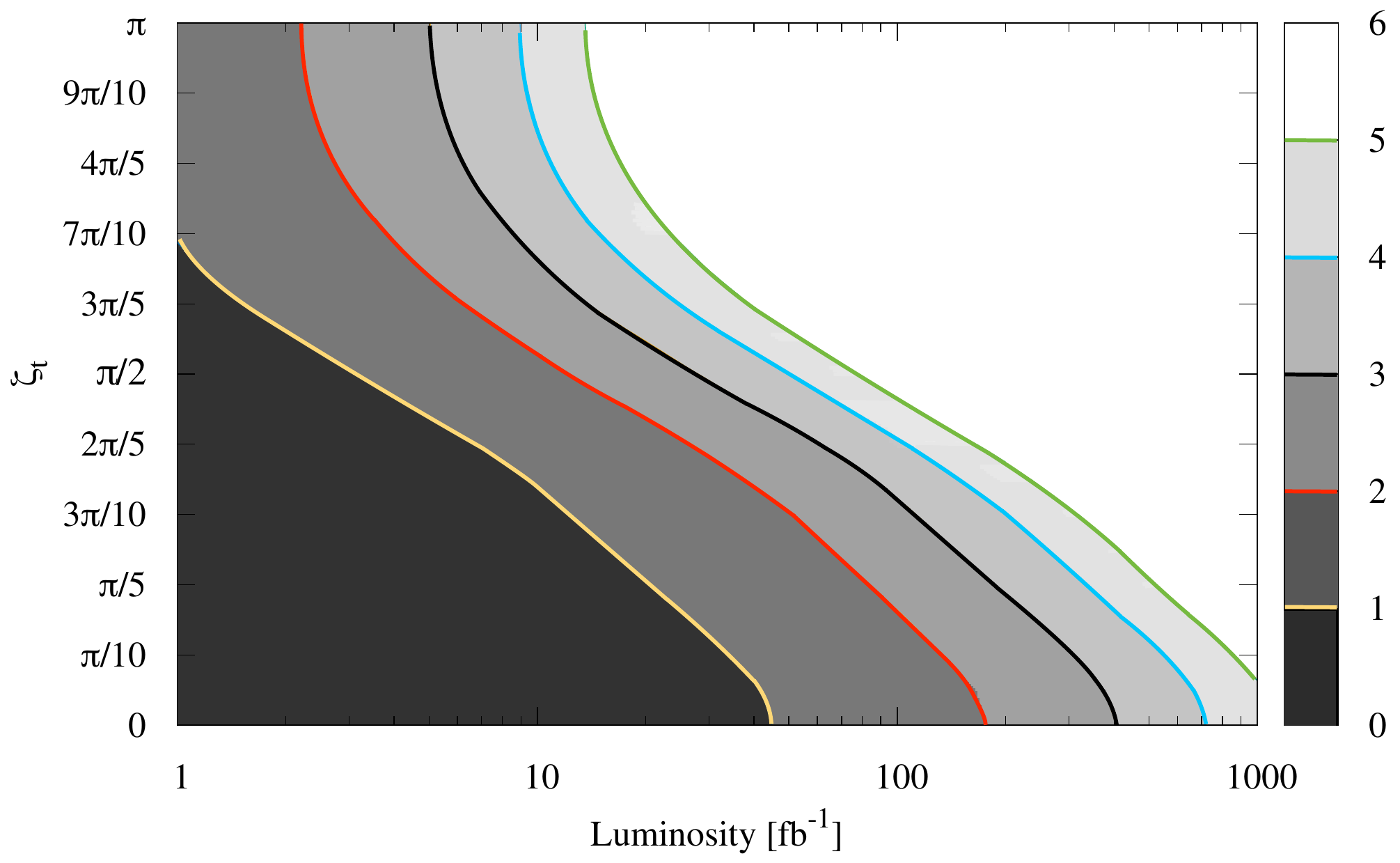}\\
\end{tabular}
\vspace{-2mm}
\caption{\small{The exclusion contour with respect to integrated luminosities at various $\zeta_t$
by considering significance based on fiducial cross section (defined in text) for $E_e =60$~GeV 
and $E_p = 7$~TeV. The regions beyond each contours are excluded for the particular luminosity, 
$black$  and $red$ solid lines correspond to $3\sigma$ and $2\sigma$ regions.}}
\label{fig:Excl2}
\end{center}
\end{figure}
In  \cref{asymm} we observed that asymmetry observables based on differential distributions of 
$\cos\theta_{b_1l^-}$ and $\Delta\phi_{b_1l^-}$ show distinct features in terms of shape although
quantitatively not very sensitive. Therefore we construct another asymmetry observable by considering the polar
angle between the sub-leading $b$-tagged jet and the lepton from $W^-$ decay, i.e, $\cos\theta_{b_2l^-}$ which
is comparatively more sensitive (quantitatively). In \cref{fig:Excl1}, we show the asymmetry $A_{\cos\theta_{b_2l^-}}$
as a function of $\zeta_t$ for $E_e = 60$ and $120$~GeV with $E_p = 7$~TeV. The
statistical uncertainties are calculated using the formula in \cref{error} for $\zeta_t = 0$ and explicitly given as:
\begin{equation}
\delta A_{\cos\theta_{b_2l^-}} = \sqrt{\frac{1 - (A_{\cos\theta_{b_2l^-}}^{\rm SM})^2 }{\sigma_{\rm SM} \cdot L}},
\label{errorsm}
\end{equation}
where $\sigma_{\rm SM}$ is total cross section of the SM signal and $A_{\cos\theta_{b_2l^-}}^{\rm SM}$ is 
numerical value of corresponding SM asymmetry. Therefore at the luminosity of $L = 1\,\rm ab^{-1}$, 
$A_{\cos\theta_{b_2l^-}}$ used to determine $\zeta_t$ within $\pi/3$ and $3\pi/5$ ($\pi/6$ and $3\pi/10$) 
at $1\sigma$ and $2\sigma$ C.L. respectively for $E_e = 60 \,(120)$~GeV.
This indicates that at low $L$ the sensitivity tends to be poorer than this, so next we use fiducial inclusive cross 
sections as another observable to find the exclusion limits.

Based on selection criteria of signal and background events discussed in  \cref{optim}, we
estimated the exclusion regions of $\zeta_t$ as a function of $L$ in fb$^{-1}$. The exclusion
is based on significance using the Poisson formula ${\cal S} = \sqrt{2[ (S+B) {\rm log}(1 + S/B) - S]}$, 
where $S$ and $B$ are the number of expected signal and background events at a particular luminosity 
respectively. Here we used 10\% systematic uncertainty for background yields only.
In \cref{fig:Excl2}, we present exclusion contours at various confidence levels for $E_e = 60$~GeV
-- understandably, higher $\sigma$-contours demand larger luminosities. It is also seen that  there is a 
kink around $\zeta_t=\pi/2$ such that for the region $0<\zeta_t<\pi/2$, we need larger luminosities for exclusion. 
This is in keeping with the feature exhibited in Fig.~\ref{fig:mu} where the constructive interference between 
the signal diagrams enhances the cross-section over the SM value much more for $\zeta_t>\pi/2$ thus requiring 
less luminosity to probe that region.
For $L = 100$~fb$^{-1}$, regions above $\pi/5 < \zeta_t \leq \pi$ and $3\pi/10 < \zeta_t \leq \pi$ 
are excluded at 2$\sigma$ and 3$\sigma$~C.L. 
While around $L = 400$~fb$^{-1}$, regions above $\pi/6 < \zeta_t \leq \pi$ and $\pi/4 < \zeta_t \leq \pi$ 
are excluded at 4$\sigma$ and 5$\sigma$~C.L. respectively.  

For higher $E_e$ = 120~GeV, the cross section for signal (background) is enhanced approximately by a factor of 
4 (3) and hence the luminosity required for exclusion is smaller compared to the $E_e = 60$~GeV case. 
Specifically, at $L = 100$~fb$^{-1}$ regions above $\pi/20 < \zeta_t \leq \pi$ and $\pi/6 < \zeta_t \leq \pi$ 
are excluded at 4$\sigma$ and 5$\sigma$~C.L. We note, as a measure of comparison, that asymmetry studies at 
the HL-LHC ~\citep{Rindani:2016scj} help probe up to $\zeta_t=\pi/6$ for a total integrated luminosity of 3~ab$^{-1}$.  
Thus, it is clear that the LHeC provides a better environment to test the CP nature of Higgs boson couplings.

Hence it is apparent that the method based on fiducial inclusive cross sections results in better limits than the 
asymmetry observable. 
It is interesting to note that for the design luminosity $L = 1\,{\rm ab}^{-1}$, almost all values of $\zeta_t$ are
excluded up to 4$\sigma$ C.L. 
While investigating the overall sensitivity of $\zeta_t$ by applying these two observables, it is also important 
to measure the accuracy of SM $tth$ coupling $\kappa$ at the LHeC energies. 
To measure the accuracy of $\kappa$ by using signal and background yields we use the formula 
${\cal K} = \sqrt{(S + B)}/{(2 S)}$ at a particular luminosity. And for $E_e = 60 \,(120)$~GeV, the measured
accuracy at the design luminosity $L = 1\,{\rm ab}^{-1}$ is given to be $\kappa = 1.00 \pm 0.17 \,(0.08)$ of 
its expected SM value, where a 10\% systematic uncertainty is been taken in background yields only.  

\section{Summary and conclusions}
\label{conc}
The discovery of a Higgs with properties very close to that predicted in the SM has necessitated
 experiments that help us elucidate the nature of its couplings. While any deviation in Higgs boson 
 couplings to $WW$ and $ZZ$ would unambiguously provide clues for a modified electroweak symmetry 
 breaking sector, any possible pseudoscalar admixture in the physical Higgs boson is more easily manifest 
 in its couplings to fermions. One promising avenue is the elucidation of such modifications in the $tth$ 
 coupling - owing to the large Yukawa, this is the most obvious channel. While the LHC is a top factory, 
 coupling determination in $pp$ colliders is usually fraught with difficulty. The $e^+e^-$ machine provides 
 a cleaner environment but one generally has to contend with smaller cross-sections. A third possibility is 
 an $e^- \, p$ machine - while this does not compete with the LHC in terms of absolute cross-sections, the
 intrinsic asymmetric nature of the machine (because of the difference in the $e^-$ and $p$ energies) 
 provides certain advantages. In this letter, we analysed the question of uncovering possible CP-odd 
 components in the $tth$ coupling at the LHeC.

Using the associated top-Higgs production and based on different observables as a function of CP-phase 
$\zeta_t$ of $tth$-coupling, we observe different distinguishable features. The difference between rapidities 
of anti-top quark and Higgs-boson $\Delta y_{\rm ht}$, and anti-top polarisation $P_t$ show unique features 
that are distinct from the pure scalar type couplings.   
 
Considering the leptonic decay mode of the anti-top quark and $h\to b \bar b$ ,we constructed the asymmetry 
observables $\Delta\phi_{b_1 l^-}$ and $\cos\theta_{b_1 l^-}$. We find that while these show deviations from 
the SM case in the region $0<\zeta_t\leq\pi/2$, the curves flatten out beyond that point. This prompted us to 
construct yet another observable $\cos\theta_{b_2l^-}$ whose variation with $\zeta_t$ is significant in the 
entire range $0<\zeta_t\leq\pi$.

Somewhat counterintuitively, exclusion regions for $\zeta_t$ obtained through fiducial cross section 
considerations result in better limits than those using asymmetry measurements. Quite strikingly, we 
find that almost all values of $\zeta_t$ can be excluded at 2$\sigma$ (4$\sigma$) with an integrated
luminosity of 200~fb$^{-1}$ (700~fb$^{-1}$) - these limits are superior to those found in studies at the 
HL-LHC. While the limits would possibly worsen when one does a full detector level simulation, our 
analysis gives excellent early signs for  the efficacy of the LHeC for coupling measurements.
 
We conclude that a study of cross-section measurements combined with accurate measurements of 
kinematic observables can be a powerful probe at the LHeC to uncover the finer details of the nature 
of the top-Higgs coupling and hope that this study adds to the physics goals of future $e^- \,p$ colliders.

As mentioned in \cref{lag}, apart from $tth$ coupling the process considered in this study involves 
$hWW$ and $Wtb$ couplings as well where non-standard anomalous contributions are not
negligible - these are studied in Refs.~\citep{Biswal:2012mp,Kumar:2015kca} 
and~\citep{Dutta:2013mva} respectively.
Since the gauge-scalar ($WWh$) and gauge-fermion ($Wtb$) anomalous couplings involve 
momentum dependent couplings, the differential distribution of final state particles is 
affected differently via such effects and can thus be used as an effective discriminant to disentangle 
the effects of different new physics contributions to the process under investigation.
For future studies, a global analysis involving all anomalous non-standard couplings together 
will be helpful to investigate the potential of precision measurement capabilities of collider
facilities like the LHeC.

\section*{Acknowledgements}

BC would like to acknowledge the support by the Department of Science and Technology under Grant 
YSS/2015/001771 and by the IIT-Gandhinagar Grant IP/IITGN/PHY/BC/201415-16. MK would like to 
acknowledge the hospitality of Indian Institute of Technology, Gandhinagar, India during the 
collaboration and SK acknowledges financial support from the Department of Science and Technology, India, 
under the National Post-doctoral Fellowship programme, Grant No. PDF/2015/000167. 
We also thank Xifeng Ruan, Claire Gwenlan for discussions while writing this article and fruitful 
discussions within the LHeC-Higgs-Top group meetings. 

\section*{References}

\bibliography{mybibfile}

\begin{thebibliography}{10}
\expandafter\ifx\csname url\endcsname\relax
  \def\url#1{\texttt{#1}}\fi
\expandafter\ifx\csname urlprefix\endcsname\relax\def\urlprefix{URL }\fi
\expandafter\ifx\csname href\endcsname\relax
  \def\href#1#2{#2} \def\path#1{#1}\fi

\bibitem{Aad:2012tfa}
G.~Aad, et~al., {Observation of a new particle in the search for the Standard
  Model Higgs boson with the ATLAS detector at the LHC}, Phys. Lett. B716
  (2012) 1--29.
\newblock \href {http://arxiv.org/abs/1207.7214} {\path{arXiv:1207.7214}},
  \href {http://dx.doi.org/10.1016/j.physletb.2012.08.020}
  {\path{doi:10.1016/j.physletb.2012.08.020}}.

\bibitem{Chatrchyan:2012xdj}
S.~Chatrchyan, et~al., {Observation of a new boson at a mass of 125 GeV with
  the CMS experiment at the LHC}, Phys. Lett. B716 (2012) 30--61.
\newblock \href {http://arxiv.org/abs/1207.7235} {\path{arXiv:1207.7235}},
  \href {http://dx.doi.org/10.1016/j.physletb.2012.08.021}
  {\path{doi:10.1016/j.physletb.2012.08.021}}.

\bibitem{Aad:2014aba}
G.~Aad, et~al., {Measurement of the Higgs boson mass from the $H\rightarrow
  \gamma\gamma$ and $H \rightarrow ZZ^{*} \rightarrow 4\ell$ channels with the
  ATLAS detector using 25 fb$^{-1}$ of $pp$ collision data}, Phys. Rev. D90~(5)
  (2014) 052004.
\newblock \href {http://arxiv.org/abs/1406.3827} {\path{arXiv:1406.3827}},
  \href {http://dx.doi.org/10.1103/PhysRevD.90.052004}
  {\path{doi:10.1103/PhysRevD.90.052004}}.

\bibitem{Khachatryan:2014jba}
V.~Khachatryan, et~al., {Precise determination of the mass of the Higgs boson
  and tests of compatibility of its couplings with the standard model
  predictions using proton collisions at 7 and 8 $\,\text {TeV}$}, Eur. Phys.
  J. C75~(5) (2015) 212.
\newblock \href {http://arxiv.org/abs/1412.8662} {\path{arXiv:1412.8662}},
  \href {http://dx.doi.org/10.1140/epjc/s10052-015-3351-7}
  {\path{doi:10.1140/epjc/s10052-015-3351-7}}.

\bibitem{Aad:2015zhl}
G.~Aad, et~al., {Combined Measurement of the Higgs Boson Mass in $pp$
  Collisions at $\sqrt{s}=7$ and 8 TeV with the ATLAS and CMS Experiments},
  Phys. Rev. Lett. 114 (2015) 191803.
\newblock \href {http://arxiv.org/abs/1503.07589} {\path{arXiv:1503.07589}},
  \href {http://dx.doi.org/10.1103/PhysRevLett.114.191803}
  {\path{doi:10.1103/PhysRevLett.114.191803}}.

\bibitem{Aad:2015mxa}
G.~Aad, et~al., {Study of the spin and parity of the Higgs boson in diboson
  decays with the ATLAS detector}, Eur. Phys. J. C75~(10) (2015) 476, [Erratum:
  Eur. Phys. J.C76,no.3,152(2016)].
\newblock \href {http://arxiv.org/abs/1506.05669} {\path{arXiv:1506.05669}},
  \href {http://dx.doi.org/10.1140/epjc/s10052-015-3685-1,
  10.1140/epjc/s10052-016-3934-y} {\path{doi:10.1140/epjc/s10052-015-3685-1,
  10.1140/epjc/s10052-016-3934-y}}.

\bibitem{Khachatryan:2014kca}
V.~Khachatryan, et~al., {Constraints on the spin-parity and anomalous HVV
  couplings of the Higgs boson in proton collisions at 7 and 8 TeV}, Phys. Rev.
  D92~(1) (2015) 012004.
\newblock \href {http://arxiv.org/abs/1411.3441} {\path{arXiv:1411.3441}},
  \href {http://dx.doi.org/10.1103/PhysRevD.92.012004}
  {\path{doi:10.1103/PhysRevD.92.012004}}.

\bibitem{AbelleiraFernandez:2012cc}
J.~L. Abelleira~Fernandez, et~al., {A Large Hadron Electron Collider at CERN:
  Report on the Physics and Design Concepts for Machine and Detector}, J. Phys.
  G39 (2012) 075001.
\newblock \href {http://arxiv.org/abs/1206.2913} {\path{arXiv:1206.2913}},
  \href {http://dx.doi.org/10.1088/0954-3899/39/7/075001}
  {\path{doi:10.1088/0954-3899/39/7/075001}}.

\bibitem{Bruening:2013bga}
O.~Bruening, M.~Klein, {The Large Hadron Electron Collider}, Mod. Phys. Lett.
  A28~(16) (2013) 1330011.
\newblock \href {http://arxiv.org/abs/1305.2090} {\path{arXiv:1305.2090}},
  \href {http://dx.doi.org/10.1142/S0217732313300115}
  {\path{doi:10.1142/S0217732313300115}}.

\bibitem{Han:2009pe}
T.~Han, B.~Mellado, {Higgs Boson Searches and the H b anti-b Coupling at the
  LHeC}, Phys. Rev. D82 (2010) 016009.
\newblock \href {http://arxiv.org/abs/0909.2460} {\path{arXiv:0909.2460}},
  \href {http://dx.doi.org/10.1103/PhysRevD.82.016009}
  {\path{doi:10.1103/PhysRevD.82.016009}}.

\bibitem{Biswal:2012mp}
S.~S. Biswal, R.~M. Godbole, B.~Mellado, S.~Raychaudhuri, {Azimuthal Angle
  Probe of Anomalous $HWW$ Couplings at a High Energy $ep$ Collider}, Phys.
  Rev. Lett. 109 (2012) 261801.
\newblock \href {http://arxiv.org/abs/1203.6285} {\path{arXiv:1203.6285}},
  \href {http://dx.doi.org/10.1103/PhysRevLett.109.261801}
  {\path{doi:10.1103/PhysRevLett.109.261801}}.

\bibitem{Kobakhidze:2016mfx}
A.~Kobakhidze, N.~Liu, L.~Wu, J.~Yue, {Implications of CP-violating Top-Higgs
  Couplings at LHC and Higgs Factories}, Phys. Rev. D95~(1) (2017) 015016.
\newblock \href {http://arxiv.org/abs/1610.06676} {\path{arXiv:1610.06676}},
  \href {http://dx.doi.org/10.1103/PhysRevD.95.015016}
  {\path{doi:10.1103/PhysRevD.95.015016}}.

\bibitem{Cirigliano:2016njn}
V.~Cirigliano, W.~Dekens, J.~de~Vries, E.~Mereghetti, {Is there room for CP
  violation in the top-Higgs sector?}, Phys. Rev. D94~(1) (2016) 016002.
\newblock \href {http://arxiv.org/abs/1603.03049} {\path{arXiv:1603.03049}},
  \href {http://dx.doi.org/10.1103/PhysRevD.94.016002}
  {\path{doi:10.1103/PhysRevD.94.016002}}.

\bibitem{Cirigliano:2016nyn}
V.~Cirigliano, W.~Dekens, J.~de~Vries, E.~Mereghetti, {Constraining the
  top-Higgs sector of the Standard Model Effective Field Theory}, Phys. Rev.
  D94~(3) (2016) 034031.
\newblock \href {http://arxiv.org/abs/1605.04311} {\path{arXiv:1605.04311}},
  \href {http://dx.doi.org/10.1103/PhysRevD.94.034031}
  {\path{doi:10.1103/PhysRevD.94.034031}}.

\bibitem{Kobakhidze:2014gqa}
A.~Kobakhidze, L.~Wu, J.~Yue, {Anomalous Top-Higgs Couplings and Top
  Polarisation in Single Top and Higgs Associated Production at the LHC}, JHEP
  10 (2014) 100.
\newblock \href {http://arxiv.org/abs/1406.1961} {\path{arXiv:1406.1961}},
  \href {http://dx.doi.org/10.1007/JHEP10(2014)100}
  {\path{doi:10.1007/JHEP10(2014)100}}.

\bibitem{Li:2017dyz}
J.~Li, Z.-g. Si, L.~Wu, J.~Yue, {Central-edge asymmetry as a probe of Higgs-top
  coupling in $t\bar{t}h$ production at LHC}\href
  {http://arxiv.org/abs/1701.00224} {\path{arXiv:1701.00224}}.

\bibitem{Rindani:2016scj}
S.~D. Rindani, P.~Sharma, A.~Shivaji, {Unraveling the CP phase of top-Higgs
  coupling in associated production at the LHC}, Phys. Lett. B761 (2016)
  25--30.
\newblock \href {http://arxiv.org/abs/1605.03806} {\path{arXiv:1605.03806}},
  \href {http://dx.doi.org/10.1016/j.physletb.2016.08.002}
  {\path{doi:10.1016/j.physletb.2016.08.002}}.

\bibitem{Kumar:2015kca}
M.~Kumar, X.~Ruan, R.~Islam, A.~S. Cornell, M.~Klein, U.~Klein, B.~Mellado,
  {Probing anomalous couplings using di-Higgs production in electron-proton
  collisions}, Phys. Lett. B764 (2017) 247--253.
\newblock \href {http://arxiv.org/abs/1509.04016} {\path{arXiv:1509.04016}},
  \href {http://dx.doi.org/10.1016/j.physletb.2016.11.039}
  {\path{doi:10.1016/j.physletb.2016.11.039}}.

\bibitem{Dutta:2013mva}
S.~Dutta, A.~Goyal, M.~Kumar, B.~Mellado, {Measuring anomalous $Wtb$ couplings
  at $e^-p$ collider}, Eur. Phys. J. C75~(12) (2015) 577.
\newblock \href {http://arxiv.org/abs/1307.1688} {\path{arXiv:1307.1688}},
  \href {http://dx.doi.org/10.1140/epjc/s10052-015-3776-z}
  {\path{doi:10.1140/epjc/s10052-015-3776-z}}.

\bibitem{Alloul:2013bka}
A.~Alloul, N.~D. Christensen, C.~Degrande, C.~Duhr, B.~Fuks, {FeynRules 2.0 - A
  complete toolbox for tree-level phenomenology}, Comput. Phys. Commun. 185
  (2014) 2250--2300.
\newblock \href {http://arxiv.org/abs/1310.1921} {\path{arXiv:1310.1921}},
  \href {http://dx.doi.org/10.1016/j.cpc.2014.04.012}
  {\path{doi:10.1016/j.cpc.2014.04.012}}.

\bibitem{Alwall:2014hca}
J.~Alwall, R.~Frederix, S.~Frixione, V.~Hirschi, F.~Maltoni, O.~Mattelaer,
  H.~S. Shao, T.~Stelzer, P.~Torrielli, M.~Zaro, {The automated computation of
  tree-level and next-to-leading order differential cross sections, and their
  matching to parton shower simulations}, JHEP 07 (2014) 079.
\newblock \href {http://arxiv.org/abs/1405.0301} {\path{arXiv:1405.0301}},
  \href {http://dx.doi.org/10.1007/JHEP07(2014)079}
  {\path{doi:10.1007/JHEP07(2014)079}}.

\bibitem{Ball:2012cx}
R.~D. Ball, et~al., {Parton distributions with LHC data}, Nucl. Phys. B867
  (2013) 244--289.
\newblock \href {http://arxiv.org/abs/1207.1303} {\path{arXiv:1207.1303}},
  \href {http://dx.doi.org/10.1016/j.nuclphysb.2012.10.003}
  {\path{doi:10.1016/j.nuclphysb.2012.10.003}}.

\bibitem{Deans:2013mha}
C.~S. Deans,
  \href{https://inspirehep.net/record/1227810/files/arXiv:1304.2781.pdf}{{Progress
  in the NNPDF global analysis}}, in: {Proceedings, 48th Rencontres de Moriond
  on QCD and High Energy Interactions: La Thuile, Italy, March 9-16, 2013},
  2013, pp. 353--356.
\newblock \href {http://arxiv.org/abs/1304.2781} {\path{arXiv:1304.2781}}.
\newline\urlprefix\url{https://inspirehep.net/record/1227810/files/arXiv:1304.2781.pdf}

\bibitem{Khachatryan:2016vau}
G.~Aad, et~al., {Measurements of the Higgs boson production and decay rates and
  constraints on its couplings from a combined ATLAS and CMS analysis of the
  LHC pp collision data at $ \sqrt{s}=7 $ and 8 TeV}, JHEP 08 (2016) 045.
\newblock \href {http://arxiv.org/abs/1606.02266} {\path{arXiv:1606.02266}},
  \href {http://dx.doi.org/10.1007/JHEP08(2016)045}
  {\path{doi:10.1007/JHEP08(2016)045}}.

\bibitem{vonBuddenbrock:2015ema}
S.~von Buddenbrock, N.~Chakrabarty, A.~S. Cornell, D.~Kar, M.~Kumar, T.~Mandal,
  B.~Mellado, B.~Mukhopadhyaya, R.~G. Reed, {The compatibility of LHC Run 1
  data with a heavy scalar of mass around 270\,GeV}\href
  {http://arxiv.org/abs/1506.00612} {\path{arXiv:1506.00612}}.

\bibitem{vonBuddenbrock:2016rmr}
S.~von Buddenbrock, N.~Chakrabarty, A.~S. Cornell, D.~Kar, M.~Kumar, T.~Mandal,
  B.~Mellado, B.~Mukhopadhyaya, R.~G. Reed, X.~Ruan, {Phenomenological
  signatures of additional scalar bosons at the LHC}, Eur. Phys. J. C76~(10)
  (2016) 580.
\newblock \href {http://arxiv.org/abs/1606.01674} {\path{arXiv:1606.01674}},
  \href {http://dx.doi.org/10.1140/epjc/s10052-016-4435-8}
  {\path{doi:10.1140/epjc/s10052-016-4435-8}}.

\bibitem{Kruse:2014pya}
A.~Kruse, A.~S. Cornell, M.~Kumar, B.~Mellado, X.~Ruan, {Probing the Higgs
  boson via vector boson fusion with single jet tagging at the LHC}, Phys. Rev.
  D91~(5) (2015) 053009.
\newblock \href {http://arxiv.org/abs/1412.4710} {\path{arXiv:1412.4710}},
  \href {http://dx.doi.org/10.1103/PhysRevD.91.053009}
  {\path{doi:10.1103/PhysRevD.91.053009}}.

\bibitem{Aaboud:2016igd}
M.~Aaboud, et~al., {Measurement of the top quark mass in the $t\bar{t}\to$
  dilepton channel from $\sqrt{s}=8$ TeV ATLAS data}, Phys. Lett. B761 (2016)
  350--371.
\newblock \href {http://arxiv.org/abs/1606.02179} {\path{arXiv:1606.02179}},
  \href {http://dx.doi.org/10.1016/j.physletb.2016.08.042}
  {\path{doi:10.1016/j.physletb.2016.08.042}}.

\bibitem{Bernreuther:2008ju}
W.~Bernreuther, {Top quark physics at the LHC}, J. Phys. G35 (2008) 083001.
\newblock \href {http://arxiv.org/abs/0805.1333} {\path{arXiv:0805.1333}},
  \href {http://dx.doi.org/10.1088/0954-3899/35/8/083001}
  {\path{doi:10.1088/0954-3899/35/8/083001}}.

\bibitem{Czarnecki:1990pe}
A.~Czarnecki, M.~Jezabek, J.~H. Kuhn, {Lepton Spectra From Decays of Polarized
  Top Quarks}, Nucl. Phys. B351 (1991) 70--80.
\newblock \href {http://dx.doi.org/10.1016/0550-3213(91)90082-9}
  {\path{doi:10.1016/0550-3213(91)90082-9}}.

\bibitem{Brandenburg:2002xr}
A.~Brandenburg, Z.~G. Si, P.~Uwer, {QCD corrected spin analyzing power of jets
  in decays of polarized top quarks}, Phys. Lett. B539 (2002) 235--241.
\newblock \href {http://arxiv.org/abs/hep-ph/0205023}
  {\path{arXiv:hep-ph/0205023}}, \href
  {http://dx.doi.org/10.1016/S0370-2693(02)02098-1}
  {\path{doi:10.1016/S0370-2693(02)02098-1}}.

\end{thebibliography}

\end{document}